\documentclass[%
 reprint,,longbibliography,preprintnumbers,
nofootinbib,
 amsmath,amssymb,
 aps,
]{revtex4-1}

\usepackage{graphicx}% Include figure files
\usepackage{dcolumn}% Align table columns on decimal point
\usepackage{bm}% bold math
\usepackage{overpic}
\usepackage[english]{babel}
\usepackage{graphicx}
\usepackage{epsfig}
\usepackage{amssymb}
\usepackage{amsmath}
\usepackage[plainpages=false,pagebackref=false,pdftex]{hyperref}
\usepackage{ulem}
\usepackage[T1]{fontenc} % if needed
\usepackage[utf8]{inputenc}
\usepackage{color}

\usepackage[titletoc]{appendix}

\begin{document}
%\begin{sloppypar}
%date{}
\title{Dynamics of chiral phase transition in a $N_f=2+1$ soft-wall AdS/QCD model}

\author{Han Tang$^1$}

\author{Bin Dong$^2$}

\author{Nanxiang Wen$^1$}

\author{Yidian Chen$^3$}

\author{Danning Li$^1$}
 \vspace{1cm}
%\email{Corresponding author: lidanning@jnu.edu.cn}

\affiliation{$^1$ Department of Physics and Siyuan Laboratory, Jinan University, Guangzhou 510632, China}

\affiliation{$^2$ Institute of semiconductors, Guangdong Academy of sciences}

\affiliation{$^3$ School of Physics, Hangzhou Normal University, Hangzhou 311121, China}

\begin{abstract}
We investigate the real-time dynamics of the chiral phase transition in a soft-wall AdS/QCD model, of which the mass plane phase diagram from equilibrium calculation is qualitatively consistent with the so-called Columbia plot. By directly solving the non-equilibrium evolution of the order parameter of the chiral phase transition, i.e. the chiral condensate, we study the thermalization of the QCD matter in different regions of the quark mass plane. It is shown that, when the system is close to the transition region, the thermalization process will show non-trivial behavior in the intermediate time region.
\end{abstract}


\maketitle

\section{Introduction}\label{introduction}
It is of both fundamental and practical importance, while being challenging, to understand the non-equilibrium dynamics of quantum many-body systems. Among the most important topics in this area, the issue of how different isolated quantum systems reach their equilibrium states, i.e. the thermalization, has attracted many theoretical attempts \cite{Mori:2017qhg}.

One of the most interesting progresses is the phenomenon named `prethermalization', which was first proposed in \cite{PhysRevLett.93.142002}. It describes a status on a time scale shorter than the thermalization time scale, on which the systems display different dynamical scaling behaviors from those at late time, as shown in many theoretical studies \cite{PhysRevLett.113.220401,PhysRevB.91.220302,PhysRevLett.118.135701}. More interestingly, a similar dynamical behavior on an intermediate time scale has been observed in a recent experiment, in a two-component Bose gas system \cite{PhysRevLett.115.245301}, which contributes to the new understandings of this topic, especially in electromagnetic-dominated cold-atom systems.

Similarly, the dynamics of another fundamental force, the strong interaction, can be investigated in the relativistic heavy ion collisions (RHICs)~\cite{BRAHMS:2004adc}. Though, after a time scale, the system could reach (local) thermal equilibrium and can be described by hydrodynamics (for a recent review, please refer to \cite{Romatschke:2017ejr}), the matter states in the early stages after collisions are far from equilibrium and require non-equilibrium methods.

Generally, there are two ways to deal with the highly non-equilibrium processes, based on the strength of the interaction~\cite{Berges:2020fwq}. At very weak coupling and when gluons are almost saturating in the system, one might take the `Glasma' picture of the strongly interacting matter and solve the classical fields to get the non-equilibrium evolution of the system~\cite{Lappi:2006fp,Gelis:2006dv}.

The other one is to use the holographic method, developed from the discovery of the anti-de Sitter/conformal field theory (AdS/CFT) correspondence~\cite{Maldacena:1997re,Gubser:1998bc,Witten:1998qj}. It is naturally a strong coupling method by mapping a strongly coupled gauge theory to a weakly coupled gauge theory.  One famous example of the success of holographic QCD is the prediction of the shear viscosity over entropy density $\eta/s=1/(4\pi)$ \cite{Policastro:2001yc,Buchel:2003tz,Kovtun:2004de}, which is very close to the value $\eta/s~$ obtained from hydrodynamic simulations \cite{Romatschke:2017ejr}. It helps to recognize the QCD matter after the collisions as strongly coupled matter.

The original AdS/CFT correspondence is a duality between two conformal theories. In order to apply such a method,  one has to extend it to non-conformal cases. Usually, one can consider some deformations upon the gravity side based on some phenomenological considerations, like thermodynamical quantities \cite{Gubser:2008ny,Gursoy:2007er,Gursoy:2007cb,Grefa:2021qvt,Cai:2022omk,Li:2011hp}, hadronic physics \cite{Erlich:2005qh,Karch:2006pv,deTeramond:2005su} for holographic QCD.

Actually, one can consider a more general Gauge/Gravity duality, considering the evolution of the fields in the holographic dimension as a mapping of the running of the dual operators and sources as the energy scale \cite{Adams:2012th}. It is interesting that an exact map of the flow equation of the functional renormalization group to the fifth dimension evolution of a scalar field can be derived in \cite{Gao:2022ojh}, though a systematic way to map and translate the 4D RG flow, especially at the IR region, is still under investigation. From such a point of view, it is reasonable to extend a general Gauge/Gravity dual to a more realistic case, with $N_c$ and $N_f$ finite values, by considering the model under some constraints from the data from experiments and lattice QCD simulations.

For non-equilibrium physics, the holographic method mapped the complex summation of the quantum fluctuations to classical gravity, so it is quite convenient for studying non-equilibrium problems. By solving the time evolution of the gravity systems, one can obtain the non-equilibrium dynamics of the dual field theories \cite{Chesler:2010bi}. Therefore, considering the large coupling constants at low temperatures and densities, which are mostly involved in the current experimental interests, we will take the holographic method to study the nonequilibrium dynamics of the QCD matter.

Based on the AdS/CFT correspondence, several efforts have been made in the thermalization of the super Yang-Mills theory (SYM) \cite{Chesler:2010bi,Grumiller:2008va,Heller:2011ju,Heller:2012je}. It is shown that the thermalization time $\tau$ is of the order of the inverse of temperature $1/T$ or the inverse of the cube root of the energy density $\mu^{-1/3}$, which is of the same order as the value used in hydrodynamic simulations \cite{Romatschke:2017ejr}. Furthermore, direct mimicings of collisions in SYM have been done in \cite{vanderSchee:2013pia,Rajagopal:2016uip}, and the results of light particle spectra are shown to be comparable with the ALICE experiment. Attempts towards thermalization in non-conformal cases have also been made in \cite{Critelli:2017euk,Craps:2013iaa}, and especially for the interesting non-equilibrium feature of first-order phase transition in \cite{Janik:2017ykj,Bantilan:2020pay,Attems:2019yqn}. All these studies have shown that the holographic method is powerful in dealing with non-equilibrium physics.

Besides the evolution of the energy-momentum tensor in the previous studies, it is also interesting and important to get the dynamic information of the order parameter when phase transitions are involved. Among the holographic QCD models, the hard-/soft-wall AdS/QCD models \cite{Erlich:2005qh,Karch:2006pv} do provide such a good starting point for this issue, since they are constructed based on symmetry and can be easily applied to discuss chiral phase transition. By introducing an additional scale relevant to chiral dynamics, one can describe the spontaneous chiral symmetry breaking at low temperature and the symmetry restoration at high temperature well~\cite{Chelabi:2015cwn,Chelabi:2015gpc,Chen:2018msc,Ballon-Bayona:2023zal,Rinaldi:2022dyh}. Furthermore, due to the match of the global symmetry in 4D field theory with the local gauge symmetry in 5D gravity, it is easy to map the conserved current to the 4D non-Abelian gauge field and introduce the charges~\cite{Chen:2019rez,Rodrigues:2020ndy,Lv:2018wfq,Cao:2020ske}. Besides, one can theoretically test the cases with different quark masses, and it turns out to give a phase diagram in the mass plane consistent with the lattice simulations (for example, see ref.~\cite{Chen:2018msc}).

Therefore, it is interesting to study the dynamic behavior of the chiral phase transition in the holographic soft-wall model. In our previous study~\cite{Cao:2022mep}, when considering the non-equilibrium evolution of the chiral condensate from non-equilibrium initial values, a very interesting phenomenon appears on the intermediate time scale, while correct thermalization behavior in the late time scale can be well described. The system is shown to be lingering over a non-equilibrium state for a long time before thermalization. Such kind of behavior is very similar to the `prethermalization' phenomena observed in many other systems\cite{PhysRevLett.113.220401,PhysRevB.91.220302,PhysRevLett.118.135701}, and it might be considered as a holographic version of this novel phenomenon. In this previous study, only $u,d$ flavors are considered. Since the strange quarks can also be active for thermodynamics, to be more realistic, we will extend our previous non-equilibrium study to the case with three non-degenerate flavors, i.e. $m_u=m_d\neq m_s$.

The paper is organized as follows. Besides the introduction part, we give a brief description of the soft-wall model, and its extension to the case with multiple flavors, in Sec.\ref{sec-sw-equilibrium}. Then, in Sec.\ref{sec-realtime}, we will derive the equation of motion for the dynamical case, and extract the evolution of the order parameter. Finally, a short summary will be given in Sec.\ref{sec-sum}.

\section{The chiral phase transition in the soft-wall AdS/QCD model}\label{sec-sw-equilibrium}

As mentioned above, we will study the dynamical process of chiral phase transition in the framework of the soft-wall AdS/QCD model. In this section, we will provide a brief review on this holographic QCD model, especially on its success in describing the chiral phase transition.

The original soft-wall model was proposed in Ref.~\cite{Karch:2006pv}, by adding a quadratic dilaton field $\Phi$ to the hard-wall AdS/QCD model \cite{Erlich:2005qh}, which is constructed by promoting the global $SU(2)_L\times SU(2)_R$ chiral symmetry to the five-dimensional (5D) gauge symmetry. The action of the soft-wall model in five-dimensional space-time takes the form
\begin{eqnarray}\label{SW-action}
S=-\int d^5x
\sqrt{-g}e^{-\Phi} &&\hspace{-1em}\text{Tr}[D_m X^+ D^m X+V( X)\nonumber\\ &&\hspace{-1em}+\frac{1}{4g_5^2}(F_L^2+F_R^2)].
\end{eqnarray}
Here, we denote the determinant of the metric $g_{MN}$ as $g$, the dilaton field as $\Phi$, both of which are considered as background fields in the probe limit. A matrix-valued complex scalar field $X^{\alpha\beta}$ is introduced as the dual field of the 4D operator $\bar{q}^\alpha q^\beta$, with $\alpha,\beta=u,d,s...$ indices in flavor space. The 4D left-/right-handed currents $\bar{q}_{L/R}\gamma^\mu t^a q_{L/R}$ are dual to the five-dimensional left-/right-handed gauge field $A_{L/R}^{\mu}\equiv A^{\mu,a}_{L/R} t^a$, with $t^a$ the generators of the $SU(N_f)$ group. In this work, we will extend the two-flavor study in the original soft-wall model to the three-flavor case, by replacing the $SU(2)$ generators with the Gell-Mann matrices for $SU(3)$. The covariant derivative $D_m$ is defined as $D_mX=\partial_mX-i A^L_mX+i XA^R_m$, while the field strength $F_{mn}$ is defined as $F_{mn}=\partial_m A_{n}-\partial_n A_{m}-i[A_m,A_n]$. In the above action, the trace is over the flavor indices. From the dimension of the dual operator, one should take $\Delta=3, p=0$ in the AdS/CFT prescription $M_5^2=(\Delta-p)(\Delta+p-4)$\cite{Witten:1998qj}, thus the leading term of the scalar potential $V(X^+X)$ should be $ M_5^2 X^{+}X=-3X^{+}X$(we take the AdS radius $L=1$), while terms of higher powers of $X$ can be included as well.

By taking the metric as the 5D AdS metric $ds^2=g_{\mu\nu}dx^\mu dx^\nu=\frac{1}{z^2}(dt^2-dz^2-d\vec{x}^2)$ and the dilaton field as a quadratic form $\Phi=\mu^2 z^2$ in the holographic dimension $z$, the original soft-wall model can describe the Regge behavior of the higher excitations, i.e. $m_n^2=4\mu^2n$. Inspired by this pioneering work, many groups extend it to describe the hadronic spectra quantitatively, by modifying the metric \cite{Sui:2009xe,dePaula:2008fp,Rinaldi:2017wdn,Contreras:2021epz}, the dialton field \cite{Kelley:2010mu}, the 5D scalar mass \cite{Chen:2022pgo,FolcoCapossoli:2016uns,Fang:2016nfj}, or dynamically solving the gravity background \cite{Li:2013oda,Li:2012ay}. Those studies verify the effectiveness of the soft-wall model in describing the hadronic physics.

The additional scale $\mu$ in the quadratic dilaton field is shown to be responsible for the linear confinement, which appears in the infrared (IR) region. As discussed in refs.
\cite{Chelabi:2015cwn,Chelabi:2015gpc,Chen:2018msc,Ballon-Bayona:2023zal,Rinaldi:2022dyh}, it is necessary to consider an additional scale responsible for the chiral dynamics. It can be accommodated in the dilaton field \cite{Chelabi:2015cwn,Chelabi:2015gpc,Chen:2018msc}, the 5D mass or equivalently the coupling of $\Phi$ and $X$\cite{Fang:2016nfj}. Here, we will follow \cite{Chelabi:2015cwn,Chelabi:2015gpc,Chen:2018msc}, and take a simple interpolation of the two scales in the dilaton filed as
\begin{eqnarray}\label{dilaton-int}
\Phi(z)&=&-\mu_1^2 z^2+(\mu_1^2+\mu_0^2)z^2 \tanh(\mu_2^2z^2).
\end{eqnarray}
It is easy to check that $\Phi(z)\sim \mu_0^2z^2$ when $z\rightarrow\infty$. Thus, the linear confinement in the higher excited states can be guaranteed. The value of $\mu_0$ should be around $\mu_0=0.43\rm{GeV}$, in order to describe the Regge slope of the light mesons. So we will follow this value in this work. For the rest two parameters in the dilaton profile, we will follow  \cite{Chelabi:2015gpc,Chelabi:2015cwn,Chen:2018msc} and take them as $\mu_1=0.83\rm{GeV}$ and $\mu_2=0.176\rm{GeV}$, since with those values the model gives a transition temperature $T_c$ of around $151.5\rm{MeV}$  and a vacuum value of chiral condensate of around $0.035\rm{GeV}^3$ for the two-flavor case, which is around those obtained from lattice simulations \cite{Aoki:2009sc,Aoki:2006br}. As for the scalar potential, we take it as
\begin{eqnarray}
V(X)=-3 X^+X+\lambda \text{Re}\{\text{det}[X]\}+\gamma (X^+X)^2.
\end{eqnarray}
It is interesting to see that under such a simple modification to the original soft-wall model, the chiral phase transition can be well described for the finite temperature case \cite{Chelabi:2015gpc,Chelabi:2015cwn,Chen:2018msc,Li:2016smq}, finite magnetic field case \cite{Li:2016gfn,Rodrigues:2018pep,Rodrigues:2018chh}, finite baryon number density \cite{Chen:2019rez,Rodrigues:2020ndy}, finite isospin density \cite{Lv:2018wfq,Cao:2020ske}, under rotation \cite{Chen:2022mhf}, and even for understanding the bubble nucleation from holography \cite{Chen:2022cgj}.  Therefore, we will stick to such a simple model and try to get the dynamics of the holographic chiral phase transition.

\begin{figure}[ht]
    \centering
     \begin{overpic}[width=0.45\textwidth]{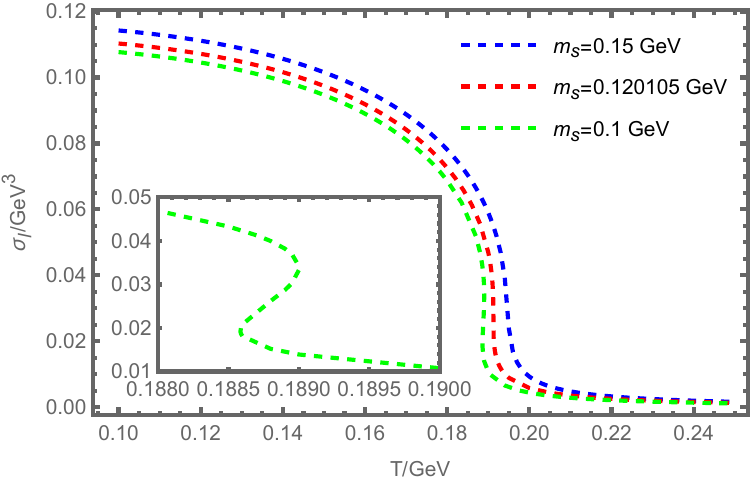}
        \put(85,20){\bf{(a)}}
    \end{overpic}
    \begin{overpic}[width=0.45\textwidth]{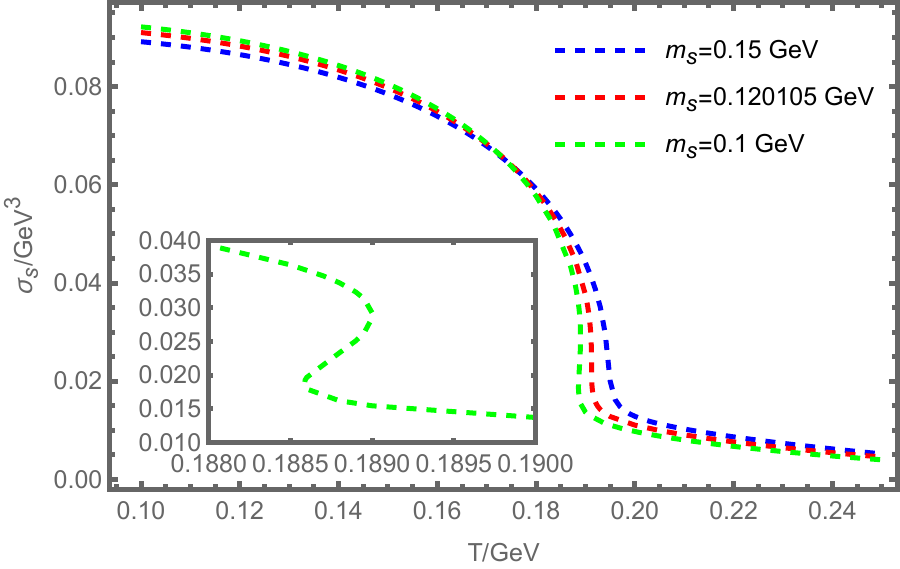}
        \put(85,20){\bf{(b)}}
    \end{overpic}
    \caption{\label{staticsigma}  Chiral condensates as functions of temperature $T$ when $m_l=0.1\rm{GeV}$ and $m_s=0.1,0.120105...,0.15\rm{GeV}$ (green, red, blue dashed lines respectively). \textbf{(a)} results for $\sigma_l$. \textbf{(b)} results for $\sigma_s$. The figures inserted are the zoom-out of the results in the range of $T=0.188-0.19\rm{GeV}$. There are three branches of solutions for $m_s=0.1\rm{GeV}$ in the region of $0.18859 \rm{GeV}<T<0.189 \rm{GeV}$.}
\end{figure}

Furthermore, to describe the finite temperature chiral phase transition, the temperature effect can be introduced by considering the black hole solutions. Following our previous work \cite{Chelabi:2015gpc,Chelabi:2015cwn}, we take the gravity background as

\begin{eqnarray}\label{bh-metric}
ds^2&=&e^{2A_s}(-f(z)dt^2-\frac{1}{f(z)}dz^2-d\vec{x}^2),\\
f(z)&=&1-\frac{z^4}{z_h^4},\\
A_s&=&-\ln(z).
\end{eqnarray}
with the horizon corresponding to the temperature through the Hawking temperature
\begin{eqnarray}
T=\frac{1}{4\pi}|f^\prime(z_h)|=\frac{1}{\pi z_h}.
\end{eqnarray}

\begin{figure}
    \centering
    \includegraphics[width=0.48\textwidth]{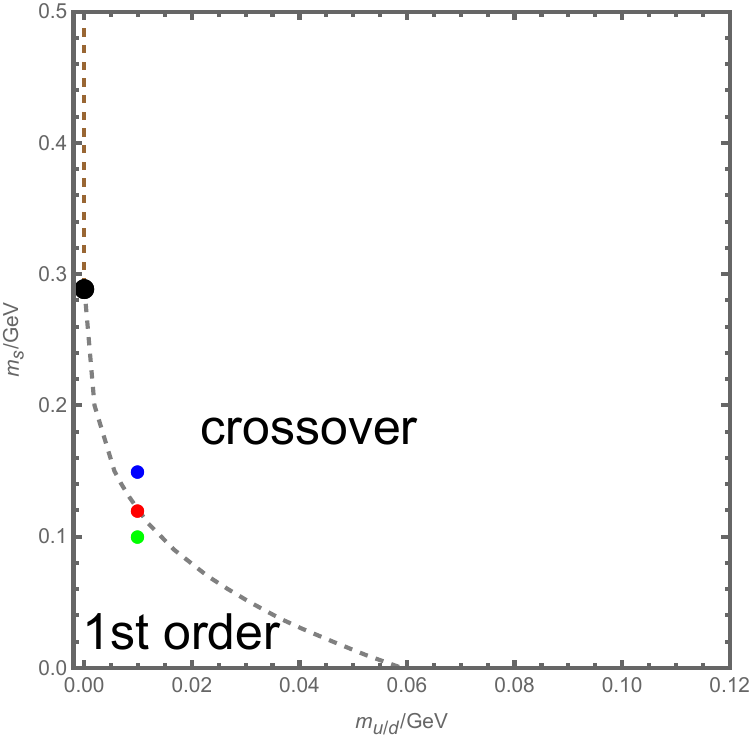}
    \caption{Phase diagram of chiral phase transition in $m_{u/d}-m_s$ plane  from the soft-wall model (please refer to \cite{Chen:2018msc} for details). Qualitatively, it is consistent with the so called `Columbia plot' from \cite{Brown:1990ev,Ding:2015ona}. The gray and brown dashed lines represent the second order transition line, which divide the whole plane into `first-order' and `crossover region'. The tri-critical point (black dot) locates at $m_l=m_u=m_d=0 \rm{GeV}, m_s=0.29\rm{GeV}$. The dynamics near the green ($m_l=0.01\rm{GeV}, m_s=0.1\rm{GeV}$), red ($m_l=0.01\rm{GeV}, m_s=0.120105...\rm{GeV}$), blue dots ($m_l=0.01\rm{GeV}, m_s=0.15\rm{GeV}$) will be studied later.}
    \label{massdiagram}
\end{figure}

Before we turn to the real-time dynamics, for the compactness of this work, we will briefly introduce the results of such a model in describing the equilibrium chiral phase transition.
Generally, since the thermodynamic impact from the heavy sectors can be effectively neglected due to their heavy mass, one can consider the light quarks $u,d,s$ only. Then, the vacuum expectation value of $X$ should follow the ansatz
\begin{eqnarray}
X=\left(
    \begin{array}{ccc}
      \frac{\chi_u(z)}{\sqrt{2}} & 0 & 0 \\
      0 & \frac{\chi_d(z)}{\sqrt{2}} & 0 \\
      0 & 0 &\frac{\chi_s(z)}{\sqrt{2}}\\
    \end{array}
  \right).
\end{eqnarray}
Here, $\chi_u,\chi_d,\chi_s$ are the expectation values of $X$, dual to $\bar{u}u, \bar{d}d,\bar{s}s$ respectively. In an equilibrium system, obviously, those expectation values are homogeneous and stable. Thus, they should depend on the holographic dimension $z$ only. In the nondegenerate case $m_u\neq m_d \neq m_s$, one will expect $\chi_u\neq\chi_d\neq\chi_s$. Then the action Eq.(\ref{SW-action}) will reduce to a simpler form as
\begin{eqnarray}
&&\hspace{-1.8em} S[\chi_u,\chi_d,\chi_s]=-\int d^5x
 \sqrt{-g}e^{-\Phi}\nonumber\\
&&\hspace{-1.8em}\times\{\sum_{\alpha=u,d,s}\left(\frac{g^{zz}}{2}\chi_\alpha^{'2}-\frac{3}{2}\chi_\alpha^2+v_4\chi_\alpha^4\right)+3v_3\chi_u\chi_d\chi_s\}.
\end{eqnarray}
Here the prime $^\prime$ denotes the derivative with respect to $z$, and $v_3\equiv\frac{\lambda}{6\sqrt{2}},v_4\equiv\frac{\gamma}{4}$ are redefinitions of $\lambda,\gamma$.
Then one can derive the equation of motion for $\chi_u,\chi_d,\chi_s$ as follows,
\begin{eqnarray}
\hspace{-1.8em}\chi_u^{''}+h(z)\chi_u^{'}+\frac{e^{2A_s}}{f}(3\chi_u-3v_3\chi_d\chi_s-4v_4\chi_u^3)&=&0,\label{eom-chiu}\\
\hspace{-1.8em}\chi_d^{''}+h(z)\chi_d^{'}+\frac{e^{2A_s}}{f}(3\chi_d-3v_3\chi_u\chi_s-4v_4\chi_d^3)&=&0,\label{eom-chid}\\
\hspace{-1.8em}\chi_s^{''}+h(z)\chi_s^{'}+\frac{e^{2A_s}}{f}(3\chi_s-3v_3\chi_u\chi_d-4v_4\chi_s^3)&=&0,\label{eom-chis0}
\end{eqnarray}
with $h(z)=3A_s^{'}-\Phi^{'}+\frac{f^{'}}{f}$.

Since both the $u$ and $d$ masses are much lighter than the strange quark, one can approximately neglect their differences, i.e. $m_u=m_d\neq m_s$. That is the $N_f=2+1$ case. In such a case, one has $\chi_l\equiv\chi_u=\chi_d\neq\chi_s$; therefore, the above equations become
\begin{eqnarray}
\hspace{-1.8em}\chi_l^{''}+h(z)\chi_l^{'}+\frac{e^{2A_s}}{f}(3\chi_l-3v_3\chi_l\chi_s-4v_4\chi_l^3)&=&0,\label{eom-chil}\\
\hspace{-1.8em}\chi_s^{''}+h(z)\chi_s^{'}+\frac{e^{2A_s}}{f}(3\chi_s-3v_3\chi_l^2-4v_4\chi_s^3)&=&0.\label{eom-chis1}
\end{eqnarray}
The ultraviolet (UV, near $z=0$ boundary) expansion of those equations can be obtained as
\begin{eqnarray}
\chi_l&=&m_l \zeta z+...+\frac{\sigma_l}{\zeta} z^3+...,\label{chiluv}\\
\chi_s&=&m_s\zeta z+...+\frac{\sigma_s}{\zeta} z^3+... ,\label{chisuv}
\end{eqnarray}
with $m_l, m_s, \sigma_l, \sigma_s$ integral constants at UV boundary, which can be mapped to $u/d$ quark mass, strange quark mass, and the chiral condensates ($\sigma_l\equiv\langle\bar{u}u(\bar{d}d)\rangle,\sigma_s\equiv\langle\bar{s}s\rangle$).
Here we follow \cite{Cherman:2008eh} and introduce a normalization constant $\zeta=\frac{\sqrt{3}}{2\pi}$ by matching the two point correlation function of $\langle\bar{\psi}\psi(q)\bar{\psi}\psi(0)\rangle$ to the 4D calculation.

Finally, when $m_s$ is larger than a critical value, the light and the strange flavor will effectively be decoupled \cite{Chen:2018msc}. Physically, this is reasonable, since the contribution to the thermodynamics becomes less when a particle becomes heavier. In this case, we need to deal with the two degenerate light flavors only. The equation of motion becomes
\begin{eqnarray}
\chi^{''}+h(z)\chi^{'}+\frac{e^{2A_s}}{f}(3\chi-4v_4\chi^3)=0.\label{eom-chi2}
\end{eqnarray}
The UV expansion can also be derived as
\begin{eqnarray}
\chi^{''}=m \zeta z+...\frac{\sigma}{\zeta} z^3+....\label{exp-chi2}
\end{eqnarray}

In this work, we will mainly stick to the $N_f=2+1$ case, with $m_u=m_d\neq m_s$. As in \cite{Chen:2018msc}, we take  $v_3=-3,v_4=8$. Then, taking a given group of the quark masses and requiring the regularity of those field configurations, one can solve the $\chi$ solutions as well as the condensations $\sigma_l,\sigma_s$ by using the `shooting method \footnote{Due to the stiffness of the equation, it is better to match solutions from both sides at an intermediate point, where the solutions from both sides can easily reach. It is not difficult to numerically check the independence on the location of this point. For details, please refer to our previous work \cite{Chen:2018msc}.}. We take $m_l=0.01\rm{GeV}$ and $m_s=0.1,0.120105...,0.15\rm{GeV}$, and show the results in Fig.\ref{staticsigma}. There, we can see that, with all these groups of quark masses, the condensates decrease from larger values at low temperature to be almost zero at high temperatures, showing phase transitions. More specifically, at $m_l=0.1\rm{GeV},m_s=0.1\rm{GeV}$, multiple branches of solutions for both $\sigma_l,\sigma_s$ appear in the region of $0.18859\rm{GeV}<T<0.189\rm{GeV}$ (see the zoom-out region in the green curves), showing a typical type of first-order phase transition. The region of the multiple solutions shrinks with the growth of $m_s$ when fixing $m_l$. At $m_s=0.120105...$, the region critically disappears and the transition becomes a second-order one, as can be seen in the red solid lines in Fig.\ref{staticsigma}. At this point, the derivative of the condensate with respect to temperature $\frac{d\sigma_{s/l}}{dT}$ diverges at $T_c\approx 0.19123...\rm{GeV}$. Then, at larger values of $m_s$, the transition becomes a smooth interpolation of the symmetry breaking and restoration phase. In this region, the divergence of $\frac{d\sigma_{s/l}}{dT}$ becomes a peak and the transition is a crossover. Usually, one can roughly define the transition at the peak location of  $\frac{d\sigma_{s/l}}{dT}$.

Then, we change the value of $m_l$, and for each $m_l$ we can obtain a critical value of $m_s^c(m_l)$. When $m_s<m_s^c(m_l)$ the transition is first-order type, while it turns to crossover when $m_s>m_s^c(m_l)$. At the critical point the phase transition is second-order type. Plotting all these critical points at the $m_l-m_s$ plane, one can get the mass diagram as shown in Fig. \ref{massdiagram}.  It is interesting to see that such a qualitative behavior of the phase diagram in the mass plane is in good agreement with the so-called `Columbia plot' \cite{Brown:1990ev,Ding:2015ona} summarized from lattice simulations and other effective studies. In this model, we get the tri-critical point at $m_l=0, m_s=0.29\rm{GeV}$. The critical lines above and below this point are shown to give different critical exponents (please refer to \cite{Chen:2018msc} for details). Since the upper segment (the gray dashed line in Fig.  \ref{massdiagram}) belongs to the same universality class as the two-flavor chiral limit, which has been studied in our previous work \cite{Cao:2022mep}, we will stick to the lower segment (the gray dashed line in Fig.  \ref{massdiagram}). In the mass diagram Fig.  \ref{massdiagram}, we also mark three points (green, red, blue dots, corresponding to the cases with the same color in Fig.\ref{staticsigma}) in different transition regions. It could be interesting to study the thermalization process in these cases.

\section{The real-time dynamics of the chiral phase transition in the soft-wall AdS/QCD model}\label{sec-realtime}

As discussed in the above section, the soft-wall model indeed gives a good qualitative description of the chiral phase diagram in the quark-mass plane. Since the previous study is based on equilibrium study on phase transitions, the dynamic information cannot be obtained. As discussed above, the real-time evolution of the thermalization process and the phase transitions might contain interesting physics. Therefore, we will try to obtain the dynamical behavior of QCD chiral phase transition in this section.

In a nonequilibrium process, most of the relevant quantities are time dependent. So one cannot take the static ansatz of $\chi$. Instead, it should depend on time as well, i.e. being $\chi(t,z)$. To reduce the numerical difficulties and focus on the time evolution, here we only turn on the time direction and leave the interplay with the spatial dimensions to the future. Under those considerations, the effective action for the relevant degrees of freedom then becomes
\begin{eqnarray}\label{eff-action}
&&S[\chi_u,\chi_d,\chi_s]=-\int d^5x
 \sqrt{-g}e^{-\Phi}\nonumber\\
&&\times\{\sum_{\alpha=u,d,s}\left(\frac{g^{tt}}{2}\dot{\chi}_\alpha^{2}+\frac{g^{zz}}{2}\chi_\alpha^{'2}-\frac{3}{2}\chi_\alpha^2+v_4\chi_\alpha^4\right)\nonumber\\
&&+3v_3\chi_u\chi_d\chi_s\}.
\end{eqnarray}
Here, the dot in $\dot{\chi}$ represents the time derivative of the fields.

We follow our previous study \cite{Cao:2022mep} and take the probe limit. Physically, it means that we have a heat bath from thermal gluonic degrees of freedom and we put the matter from the flavor sector into the bath. Then, we try to study the real-time dynamics of the flavor matter. In this way, the metric will be taken as the static one in eq.\eqref{bh-metric}. Since it is not easy to deal with the apparent singular behavior near $z_h$ with the Poincare coordinate,  it is more convenient to use the Eddington-Finkelstein (EF) coordinates, by taking the following transformation
\begin{eqnarray}\label{trans}
\nu=t-h(z),
\end{eqnarray}
with $h^{\prime}(z)=\frac{1}{f(z)}$. Under this transformation, the metric becomes
\begin{eqnarray}\label{EFmetric}
ds^2=e^{2A(z)}\{f(z) d\nu^2+2 d\nu dz-d\vec{x}^2\}.
\end{eqnarray}
Then, one can easily derive the time evolution of $\chi$s as shown in eqs. (\ref{eom-EF-time-chil},\ref{eom-EF-time-chis}).

\begin{widetext}
\begin{eqnarray}
2\partial_\nu\partial_z \chi_l-\left[\frac{3}{z}+\Phi '(z)\right]\partial_\nu \chi_l -f(z)\partial_z^2\chi_l&\nonumber\\
+\left[\frac{3}{z} f(z)+\Phi'(z)f(z)-f'(z)\right]\partial_z\chi_l+\frac{1}{ z^2}(3-3v_3\chi_s-4v_4\chi_l^2)\chi_l&=0.\label{eom-EF-time-chil}\\
2\partial_\nu\partial_z \chi_s-\left[\frac{3}{z}+\Phi '(z)\right]\partial_\nu \chi_s -f(z)\partial_z^2\chi_s&\nonumber\\
+\left[\frac{3}{z} f(z)+\Phi'(z)f(z)-f'(z)\right]\partial_z\chi_s+\frac{1}{ z^2}(3\chi_s-3v_3\chi_l^2-4v_4\chi_s^3)&=0.\label{eom-EF-time-chis}
\end{eqnarray}
\end{widetext}

It is obvious that the order of time derivative reduces to first order. So, it becomes easier to solve the initial problems by using the pseudospectral method \footnote{For details, please refer to \cite{boyd2001chebyshev} or the appendix in \cite{Cao:2022mep}. }. One can also get similar expansions with eqs.(\ref{eom-EF-time-chil},\ref{eom-EF-time-chis}), but with time dependent condensates $\sigma_l(\nu),\sigma_s(\nu)$. So, after one numerically solves the time dependence of $\chi$s, one can obtain the time dependence of $\sigma_l(\nu),\sigma_s(\nu)$. Here, though the transformation from $t$ to $\nu$ is implicit, one can verify that at the boundary $z=0$ the two are the same. Thus, we can conveniently obtain the $t$ dependence by directly solving eqs.(\ref{chiluv},\ref{chisuv}). Before one solves those equations, the initial states should be prepared in some ways. In \cite{Cao:2022mep}, we solve the equilibrium states at some initial temperatures $T_i$ and take them as the initial states. Then, we suddenly quench the system to another temperature by setting the values of $z_h$. However, since we are interested in the qualitative behavior near the transition temperatures, it could be a good approximation to take the initial states as $\chi_l(\nu=0)=m_l\zeta z+...+\lambda_l\pi^3 T^3 z^3,\chi_s(\nu=0)=m_l\zeta z+...+\lambda_s\pi^3 T^3 z^3$ with the dimensionless parameter $\lambda_l,\lambda_s$ free parameters to determine the initial states.

\subsection{The 1st order region}

\begin{figure}[ht]
    \centering
     \begin{overpic}[width=0.45\textwidth]{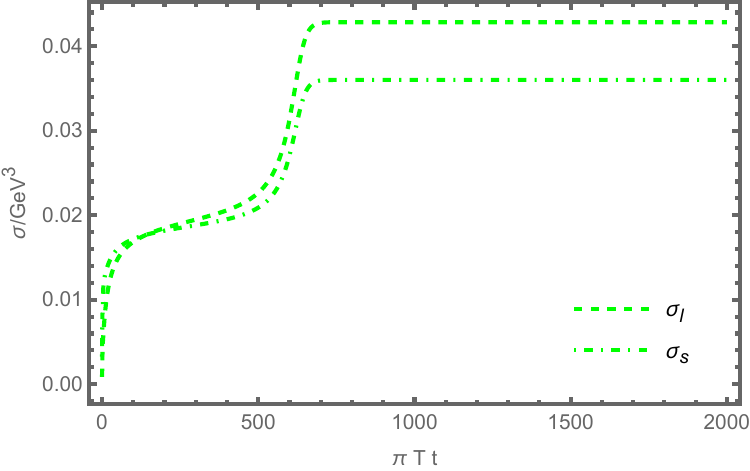}
        \put(85,30){\bf{(a)}}
    \end{overpic}
    \begin{overpic}[width=0.45\textwidth]{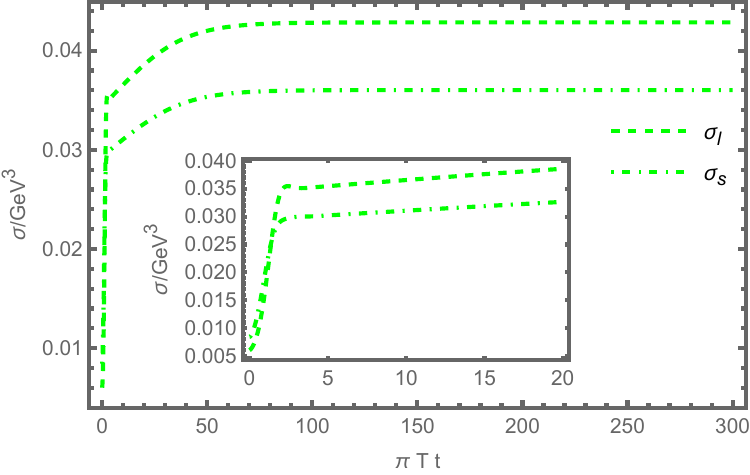}
        \put(85,30){\bf{(b)}}
    \end{overpic}
    \caption{\label{firstorderrelaxation}  Thermalization with different initial conditions at $m_l=0.01\rm{GeV}, m_s=0.1\rm{GeV}$ in the first order region . The quenched temperature is taken as $T=0.18855\rm{GeV}$, close to the transition region ( with the multiple solutions).\textbf{(a)} $\lambda_l=\lambda_s=0.01$. \textbf{(b)} $\lambda_l=\lambda_s=0.1$.}
\end{figure}

In this section, we will consider the thermalization in the first order region. We take $m_l=0.010\rm{GeV}, m_s=0.1\rm{GeV}$ as an example. As discussed above, we will simply take the initial states as $\chi_l(\nu=0)=m_l\zeta z+...+\lambda_l\pi^3 T^3 z^3,\chi_s(\nu=0)=m_l\zeta z+...+\lambda_s\pi^3 T^3 z^3$. Obviously, those states are not in equilibrium, and the system will start from these states and relax to its equilibrium states. Firstly, we consider the quenched temperature to be $T=0.18855\rm{GeV}$, near the phase transition region (with multiple solutions). Then we take $\lambda_l=\lambda_s=0.01$, which gives small initial values of condensates. By numerically solving eqs.(\ref{eom-EF-time-chil},\ref{eom-EF-time-chis}), we obtain the time evolution of
condensates as shown in Fig.\ref{firstorderrelaxation} (a). There we can see that after a long time $t>700\frac{1}{\pi T}$, both $\sigma_l$ and $\sigma_s$ reach their equilibrium values. At this stage, the system exponentially thermalize to its equilibrium state. More interestingly, the system does not directly thermalize to its equilibrium state. Instead, it turns to a middle state with condensates around $0.02\rm{GeV^3}$, which are almost the lower branch in the multiple-solution region, as shown in Fig.\ref{staticsigma}. It seems the system relaxes to a metastable state first and then transits to the equilibrium state. Then, we take $\lambda_{l}=\lambda_{s}=0.1$ and study the evolution of condensates. The results are shown in Fig. Fig.~\ref{firstorderrelaxation} (a). Though the system does not relax to the metastable state, it does not exponentially thermalize to its equilibrium state. Instead, in the intermediate time region, the condensates grow rapidly to some states and then slowly relax to the equilibrium state. Thus, we see that, when the quenched temperature is close to the transition region, the thermalization of the system will show some non-trivial behavior in the intermediate time.

\begin{figure}
    \centering
    \includegraphics[width=0.45\textwidth]{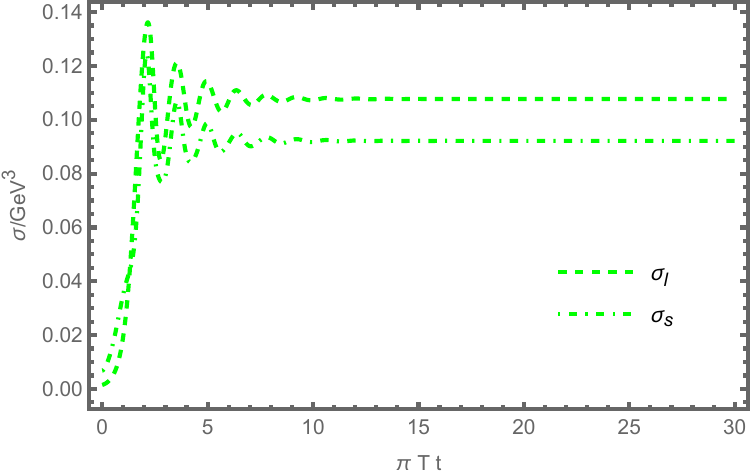}
    \caption{Thermalization with $\lambda_l=\lambda_s=0.1$} at $m_l=0.01\rm{GeV}, m_s=0.1\rm{GeV}$ in the first order region. The quenched temperature is taken as $T=0.1\rm{GeV}$, far from the transition region ( with the multiple solutions).
    \label{firstorderrelaxation-1}
\end{figure}

As a check, we also study the cases when the quenched temperature is far from the transition region. We take $T=0.1\rm{GeV}$, $\lambda_l=\lambda_s=0.1$, and show the time evolution in Fig.\ref{firstorderrelaxation-1}. There we can see that the condensates relax to their equilibrium values very quickly, at order of $\frac{1}{\pi T}$, much smaller than the values at temperatures close to the transition region. Moreover, we can observe oscillations of condensates in the intermediate region. This is because the complex frequency of the corresponding fluctuation mode has a non-zero real part (for more details, please refer to \cite{Cao:2022mep}).

\subsection{The 2nd order line}

In \cite{Cao:2022mep}, we studied the thermalization process when the quenched temperature is set at the critical point. It should be interesting to extend those studies to $N_f=2+1$. Besides belonging to different universality classes, another non-trivial difference is that the condensates at the critical point of $N_f=2$ cases are finite while for the two flavor chiral limit they are vanishing. Therefore, there could be more abundant physics in the $N_f=2+1$ case.

\begin{figure}[ht]
    \centering
     \begin{overpic}[width=0.45\textwidth]{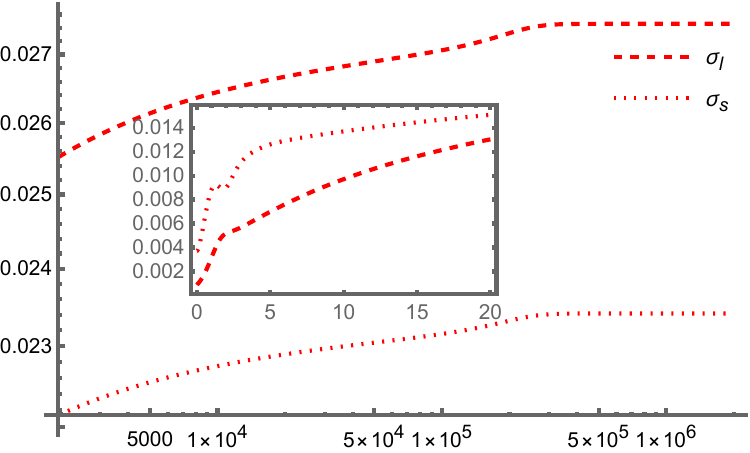}
        \put(85,30){\bf{(a)}}
    \end{overpic}
    \begin{overpic}[width=0.45\textwidth]{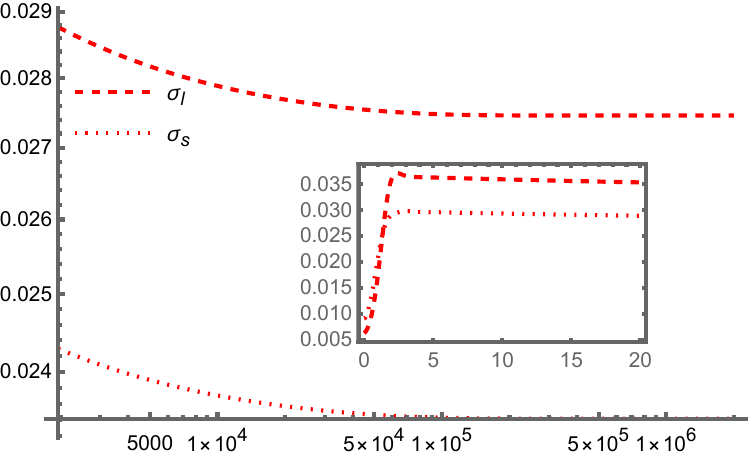}
        \put(85,49){\bf{(b)}}
    \end{overpic}
    \caption{Relaxation in critical lines. The quark masses are taken as $m_l=0.01\rm{GeV}, m_s=0.120105...\rm{GeV}$. The quenched temperature is taken as $T=T_c=0.19123...\rm{GeV}$. Due to the critical slowing down and the long relaxation time near the critical line, we use a double logarithmic plot. \label{criticalrelaxation1}  \textbf{(a)} $\lambda_l=\lambda_s=0.01$. \textbf{(b)} $\lambda_l= \lambda_s=0.1$.}
\end{figure}

We take $m_u=0.01\rm{GeV}, m_s=0.120105...\rm{GeV}$ (the red dot in Fig.\ref{massdiagram}), which locates at the lower segment in the critical boundaries. Then, we take the initial values $\lambda_l=\lambda_s=0.01$, and quenched the system to $T=T_c=0.19123...\rm{GeV}$. The evolution of the condensates is shown in Fig.\ref{criticalrelaxation1} (a). From the plot, it is shown that the system take a very long time to thermalize to the equilibrium state, and it shows the critical slowing down behavior\footnote{Different from the two-flavor case, due to the non-vanishing condensate at the second order transition point, it is hard to determine the critical temperature and critical at very high accuracy. Therefore, the relaxation time obtained is finite though very large. But if neglect the very long time tails, it is a kind of critical slowing down.}. Furthermore, in the zoom-out plot in Fig.\ref{criticalrelaxation1} (a), one can see some non-trivial behavior in the intermediate time region as well.

We start from another initial states by taking $\lambda_l=\lambda_s=0.1$. The results are shown in Fig.\ref{criticalrelaxation1} (b). In such a case, the condensates increase rapidly at a time close to the intermediate time in Fig.\ref{criticalrelaxation1} (a). Then starting from oversaturate values, the condensates decrease and relax to their equilibrium values. We also see the critical slowing down and a very long relaxation time.

\begin{figure}
    \centering
    \includegraphics[width=0.45\textwidth]{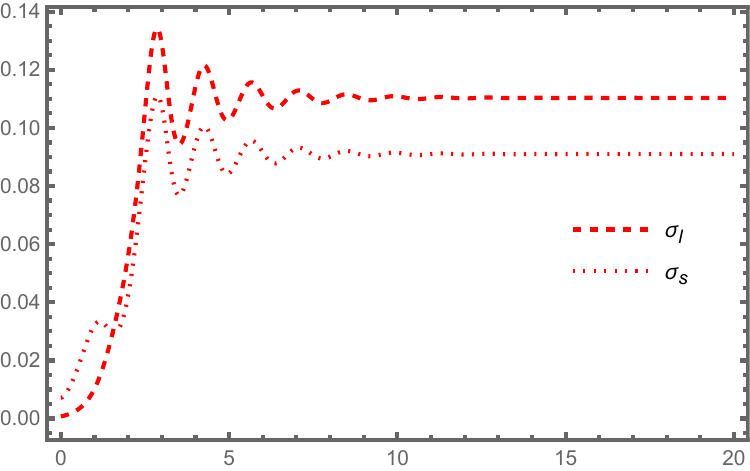}
    \caption{Thermalization with $\lambda_l=\lambda_s=0.01$} at $m_l=0.01\rm{GeV}, m_s=0.120105\rm{GeV}$ in the critical line. The quenched temperature is taken as $T=0.1\rm{GeV}$, far from the critical temperature.
    \label{relaxationfarcritical}
\end{figure}

As a comparison, we take the quenched temperature as $T=0.1\rm{GeV}$, far away from the critical temperature. It is shown in Fig.\ref{relaxationfarcritical} that the system thermalize fast to its equilibrium states.

\subsection{The crossover line}
\begin{figure}
    \centering
    \includegraphics[width=0.45\textwidth]{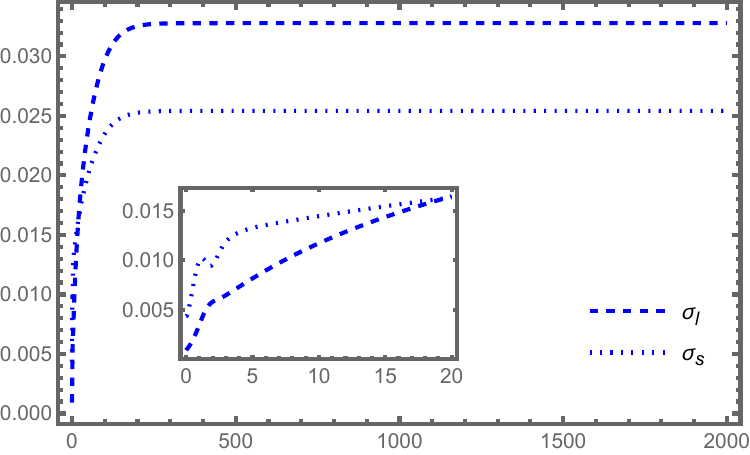}
    \caption{Thermalization with $\lambda_l=\lambda_s=0.01$} at $m_l=0.01\rm{GeV}, m_s=0.15\rm{GeV}$ in the crossover region. The quenched temperature is taken as $T_p=0.194487\rm{GeV}$, at which $-\frac{d \sigma_{l/s}}{dT}$ reach their maximums.
    \label{relaxationcrossover}
\end{figure}

Finally, we take $m_u=0.01,0.1\rm{GeV}$, which locate at the crossover region. We also determine the pseudo-critical temperature $T_p=0.194487\rm{GeV}$, at which $-\frac{d \sigma_{l}}{dT}$ and $-\frac{d \sigma_{s}}{dT}$ reach their maximums. We take the quenched temperature to be $T_p$, and let the system evolve with the initial states $\lambda_l=\lambda_s=0.01$. The results are given in Fig.\ref{relaxationcrossover}. From the results, we can see that since the divergence of $-\frac{d\sigma_{l/s}}{dT}$ becomes a peak, there is no exact critical slowing down behavior. The system relaxes quickly to the equilibrium states. However, in the intermediate region, we can still see some non-trivial behavior, as shown in the zoom-out plot in Fig.\ref{relaxationcrossover}.

\section{Conclusions and discussions}\label{sec-sum}
We extend our previous non-equilibrium holographic study in \cite{Cao:2022mep} to the case with multiple non-degenerate flavors. By extracting the real-time dynamical evolution of the order parameters, i.e. the chiral condensates $\sigma_l, \sigma_s$, we study the thermalization process from far-from-equilibrium initial states to their equilibrium states.

From the numerical results, it is shown that at long time regions, the system will thermalize to its equilibrium states, independent of the initial conditions. Near the $N_f=2+1$ critical lines, the system will also show some critical slowing down behavior, though the critical lines in $N_f=2+1$ are quite different from those in the $N_f=2$ case. More interestingly, in the intermediate region, the evolution of the system would show some non-trivial behavior when the system is quenched to near the phase transition point. However, in the current study, we focus only on the qualitative behavior, and the quantitative analysis on the dynamic scaling law is not considered, which will be left for the future. In this work, the dynamical critical behavior is investigated near the critical point without the interplay of the baryonic number densities. It is worth testing the difference when a critical end point appears with finite baryon number density.

\vspace*{1cm}

{\bf Acknowledgements}\quad
This work is supported by the National Natural Science Foundation of China under Grant Nos. 12275108, 12235016, 12305136, the GDAS' Project of Science and Technology Development, China (2020GDASYL-20200103117), and the start-up funding of Hangzhou Normal University under Grant No. 4245C50223204075.

\vspace*{1cm}

\bibliographystyle{apsrev4-1}
\bibliography{refs}
\end{document}